\documentclass[12pt,a4paper]{article}
\usepackage{graphicx}
\usepackage{amsfonts}
\usepackage{amssymb}
\usepackage{amsmath}
\usepackage{multicol}

\begin{document}

\title{1+1-dimensional Yang-Mills equations and mass via   quasiclassical  correction to action}
\author{\textbf{Sergey Leble},\\
Immanuel Kant Baltic Federal University,  
\\
Al. Nevsky st 41, Kaliningrad, \\ Russia , \\
\texttt{lebleu@mail.ru}}

\maketitle
\begin{abstract}
Two-dimensional Yang-Mills  models in a pseudo-euclidean space are considered from a
point of view of a class of nonlinear Klein-Gordon-Fock equations. It is shown that the Nahm reduction does not work, another choice is proposed and investigated.   A quasiclassical quantization of the models is based 
on Feynmann-Maslov  path integral construction  and its  zeta function representation in terms of a
Green function diagonal for an auxiliary heat equation with an elliptic
potential. The natural renormalization use a freedom in vacuum state choice as well as  the choice of the norm of an evolution   operator eigenvectors. A nonzero mass appears  via the quasiclassical correction.
 \end{abstract}
 
 \section{Introduction. On Nahm models.}
{\bf Underlying ideas} for  
this investigation, related to the classical Yang-Mills (YM) theory reductions, were taken from works of Baseyan \cite{Bas}, Corrigan \cite{Cor} and Nahm   \cite{Nahm2}. 

This paper is a direct development of author's results \cite{L} in which one-dimensional model immersed in SU(2) YM theory  was studied in the context of Nahm model. The  author's main result   \cite{L} is
  a demonstration of existence and evaluation of nonzero  quantum correction to action against classical zero enertgy (representing mass) as a consequence of the proposed model. The one-dimensional Yang-Mills-Nahm  models were considered from  algebrogeometric points of view.  A quasiclassical quantization of the models is based 
on  Maslov version of path integral construction  and its  zeta function representation in terms of a
Green function diagonal for an auxiliary heat equation with an elliptic
potential.  The Green function diagonal and, hence, the generalized zeta function and its derivative are expressed via solutions of Drach equation \cite{D} and, alternatively, by means of  Its-Matveev  \cite{IM} formalism in terms of  Riemann theta-function.  The approach is  based on Baker-Akhiezer functions
for Kadomtsev-Petviashvili equation \cite{L1}. The quantum corrections to action
  of the   model
   are evaluated. The fields from the class of elliptic functions
   are properly
   studied.       
For such model, which field is  represented via elliptic
(lemniscate) integral by construction, YM field mass
is defined as  the  quantum correction, in the quasiclassical approximation it is evaluated  via hyperelliptic
integral.

The model is related via the
(Atiyah-Drinfeld-Hitchin-Manin- Nahm) construction to static
monopole solutions to
 Yang-Mills-Higgs theories in four
dimensions in the
 Bogomolnyi-Prasad-Sommerfield
limit. 
The ADHMN construction $ \Leftrightarrow $ equivalence
between  self-dual equations, one - unidimensional, the other in
three dimensions (reduced  Euclidean four dimensional theory by
deleting dependence on a single variable), see E. Corrigan et al \cite{Cor}
.

The weak point of   description starting from the 1+0 Nahm model  is namely the one-dimensionality of the reduction that provoke  ambiguity of the interpretation of the correction as the mass.

\textbf{Yang-Mills equations} in PseudoEuclidean dimensions.
The equation for YM field $T_{\mu}$ from semisimple compact gauge group in covariant form reads as
 \begin{equation}\label{YM}
   \nabla^{\mu}T_{\mu\nu}=0,
\end{equation}
$\mu,\nu=0,1,2,3$, time variable is   $x_0 =ct, c=1, x_k$ - space variables. For the gauge fields $T_{\mu} = T_{\mu}^+$, where
 \begin{equation}\label{Tmn}
T_{\mu\nu} = \partial_{\mu}T_{\nu} -  \partial_{\nu}T_{\mu}- [T_{\mu},T_{\nu}],
\quad \partial_{\nu}\Phi = \partial_{\mu} - [T_{\mu},\Phi].
\end{equation}
one have
\begin{equation}\label{YM3}
\begin{array}{c}
	\square T_{\nu} - \partial_{\nu}\partial_{\mu}T_{\mu} +[T_{\mu},\partial_{\nu}T_{\mu}- \partial_{\mu}T_{\nu}]+[T_{\mu},[T_{\mu},T_{\nu}]]\\
	-\partial_{\mu}[T_{\mu},T_{\nu}]=0,
	 \end{array}
\end{equation}
as written in e.g. Faddeev-Slavnov book \cite{FS}.

The reduction via independence on  $x_k$, k=1,2,3;  setting $x_0 =t$, choosing the Hamilton gauge $T_0=0$, gives
\begin{equation}\label{YM1}
\begin{array}{c}
     \frac{d^2T_k}{dt^2} = [T_j[T_j,T_k]], \quad
  [T_k,\frac{d T_k}{dt}]=0. \\
\end{array}
\end{equation}
The self-dual equations \cite{Cor},
\begin{equation}\label{SDE}
    \frac{dT_i}{dt}=\pm \varepsilon_{ijk}T_jT_k,
\end{equation}
 imply Eqs. (\ref{YM1}).

 For illustration we would use 2x2 matrix gauge group  (isospin group SU(2)) and the basis of Pauli matrices $\sigma_i$ ,
expanding $T_\mu=A_{\mu}^k\sigma_k$. Equalizing terms by $\sigma_k$ and evaluating sums one goes to the vector form
\begin{equation}
\begin{array}{c}
 \square A_{\nu}^k -  \partial_{\nu}\partial_{\mu}A_{\mu}^k + A_{\mu}^j\varepsilon_{jpk}(i\partial_{\nu}A_{\mu}^p - i\partial_{\mu}A_{\nu}^p) \\
- iA_{\mu}^j A_{\mu}^jA_{\nu}^k + iA_{\mu}^jA_{\mu}^kA_{\nu}^j -i\varepsilon_{jpk}A_{\nu}^p\partial_{\mu}A_{\mu}^j-i\varepsilon_{jpk}A_{\mu}^j\partial_{\mu}A_{\nu}^p.
	 \end{array}
\end{equation}
{\bf YM equations : vector form, Lorentz gauge.}

Rescaling the vector potential we return the self-action charge parameter $\epsilon$ to rewrite the YM equation keeping the same notations 
 \begin{equation}\label{vec}
\begin{array}{c}
 \square \vec A_k +  2\epsilon \vec A^{\mu}\times (2 \partial_{\mu}\vec A_k - \partial_{k}\vec A_{\mu} -2\epsilon \vec A_k \times \vec A_{\mu})=0,
	 \end{array}
\end{equation}
$k=1,2,3, $
where, $\vec A_0$ is expressed  from the Lorentz gauge 
 \begin{equation}\label{Lor}
\vec A_{0}= \partial_{0}^{-1}\partial_{k}\vec A_{k},
\end{equation}
e.g. see Konopleva-Popov book \cite{KP}. The difference is in that we use real time variable.

The units are chosen so as velocity of light in vacuum  $c=1$, hence $\square=\partial_0^2-\partial_k\partial_k.$

Quantization is performed in Faddeev-Popov works \cite{FP} and presented in details icluding perturbation technique in 
Faddeev L.: \cite{Fad}.
Recently we evaluated correction to the mass for the Nahm reduction of YM theory  by means of quasiclassical asymptotics \cite{L,L1} developing its renormalization in \cite{KwL} with applications to the special case of Heisenberg chain equation, that differs from Nahm case only by  physical origin and rescaling.

\textbf{ Regularization (renormalization) as expalnation of nonzero mass appearance by quantization}

Faddeev: " Sidney Coleman  coined a nice name \textsl{dimensional transmutation} for
the phenomenon, which I am going to describe. Let us see what all this
means."

"Through these (free particles) solutions are introduced
via well defined quantization of the free fields. However the more thorough
approach leads to the corrections, which take into account the selfnteraction
of particles" \cite{Fad}.
 
\textbf{ The task of the present work} is the derivation and solution of the field equations for a class of the  two dimensional models (Sec. \ref{NR}). The result of the reduction of the basic YM equations and  the corresponding Lagrangian  is similar to the one-dimensional one \cite{L}: we obtain  1+1 $\phi^4$ (phi-in-quadro) model equations with the zero mass term and coefficients that depend on algebraic closure of an matrix anzatz for the gauge fields that fix the model.    
The stationary and directed waves (Sec. \ref{DW}) are thought as quasiperiodic  solutions of the model equations that  are expressed   in  terms of elliptic functions. Its quantization (Sec\ref{Q}) is again performed by means reduced Lagrangean (Sec \ref{L}) for quasiclassical Feynman-Maslov integral, which evaluation and quantum corrections to action (Sec. \ref{M}) is      
based on the mentioned technique of the generalized zeta-function renormalization in terms of  the nonlinear Drach equation (Sec. \ref{Dr}   ). It is derived  for the Green function diagonal (within the heat kernel formalism) and gives polynomial solutions in elliptic variables.

Extra variables of arbitrary dimensions (Sec. \ref{Ex}, App.) are accounted for the model   applications
   of the solutions in elementary particles  physics.

\section{The case of 1+1 dimension and reductions}

\subsection{General equations in the vector form and Nahm reduction}

In 1+1 space,  classical YM theory \cite{Boo}, Eq. \ref{YM3} with the Hamilton reduction                                                                                                                   
$T_0=0$   gives                                              
\begin{equation}
\begin{array}{c}
	\square T_{k} + \partial_{k}\partial_{s}T_{s} -[T_{s},\partial_{k}T_{s}]+2[ T_{s},\partial_{s}T_{k}] 	+[\partial_{s}T_{s},T_{k}] \\  
 -[T_{s},[T_{s},T_{k}]]	=0,
	\end{array}
\end{equation}
Nahm reduction $T_s=A\alpha_s$ simplifies it as
\begin{equation}
\begin{array}{c}
	\alpha_k\square  A  +\alpha_s \partial_{k}\partial_{s}A-[A\alpha_s,\partial_{k}A\alpha_s]+3A\partial_{s}A[\alpha_s,\alpha_k] 	 \\ 
 -A^3[\alpha_s,[\alpha_s,\alpha_k]]	=0,
	\end{array}
\end{equation}
that fails in 1+1. Namely, taking 
k=1
\begin{equation}
\begin{array}{c}
	\alpha_1\square  A  +\alpha_1 \partial_{1}\partial_{1}A-[A\alpha_1,\partial_{1}A\alpha_1]+3A\partial_{1}A[\alpha_1,\alpha_1 ]	 \\ 
 -A^3[\alpha_s,[\alpha_s,\alpha_1]]	=\alpha_1\partial_0^2A-[\alpha_s,[\alpha_s,\alpha_1]]A^3=0,
	\end{array}
\end{equation}
 one arrive at  ODE, while for  k=2 we have
\begin{equation}
\begin{array}{c}
	\alpha_2\square  A  +\alpha_1 \partial_{2}\partial_{1}A-[A\alpha_1,\partial_{2}A\alpha_1]+3A\partial_{1}A[\alpha_1,\alpha_2] 	 \\ 
 -A^3[\alpha_s,[\alpha_s,\alpha_2]]	=\alpha_2\square  A  + +3A\partial_{1}A[\alpha_1,\alpha_2]-A^3[\alpha_s,[\alpha_s,\alpha_2]]= 0,
	\end{array}
\end{equation}
that necessarily reduces to 1D case.

\subsection{Novel reduction}
\label{NR}
We  use the Lorentz gauge,  more natural for waves description  and for  the vector form ( \ref{vec}) as more transparent.
 So, let us consider alternative (compared to Nahm one) proposal of reduction: the field is specially prepared as
 \begin{equation}\label{red}
\vec A_k=\phi_k(x,t)\vec s_k
\end{equation}
where $\vec s_k$ are constant vectors in isotopic space. It may mean that a particle space state component is linked with the isotopic one.  Plugging \eqref{red}  in \eqref{vec} and returning to low indices, write
\begin{equation}
\begin{array}{c}
 \square \phi_k\vec s_k +  2\epsilon \vec A_{0}\times (2 \partial_{0}\phi_k\vec s_k - \partial_{k}\vec A_{0} - 2\epsilon\phi_k\vec s_k \times \vec A_{0})-\\
	   2\epsilon\phi_j\vec s_j\times (2 \partial_{j}\phi_k\vec s_k - \partial_{k}\phi_j\vec s_j- 2\epsilon\phi_k\vec s_k \times \phi_j\vec s_j)=0.
 \end{array}
\end{equation}
The Eq. \eqref{Lor} in 1+1 reads 
$$
\partial_{0}\vec A_{0} =\partial_{1}\phi_1\vec s_1,
$$
so, taking the Eq. (\ref{vec})
along the reduction, we write
\begin{equation}\label{bs}
\begin{array}{c}
 \square \phi_k\vec s_k + 2\epsilon \partial_{0}^{-1}\partial_{1}\phi_1\vec s_1\times (2 \partial_{0}\phi_k\vec s_k -\\ \partial_{k}\vec  \partial_{0}^{-1}\partial_{1}\phi_1\vec s_1 -2\epsilon\phi_k\vec s_k \times \partial_{0}^{-1}\partial_{1}\phi_1\vec s_1)-\\
	 2\epsilon \phi_j\vec s_j\times (2 \partial_{j}\phi_k\vec s_k - \partial_{k}\phi_j\vec s_j-2\epsilon\phi_k\vec s_k \times \phi_j\vec s_j)=0.
 \end{array}
\end{equation}
Scalar product of\eqref{bs} with $\vec s_k$  gives
  \begin{equation}
\begin{array}{c}
 \square \phi_k(\vec s_k,\vec s_k) + 4\epsilon ^2\phi_k(\partial_{0}^{-1}\partial_{1}\phi_1)^2 (\vec s_k,\vec s_1\times (   \vec s_k \times \vec s_1))+\\
	 4\epsilon^2  \phi_k\phi_j\phi_j(\vec s_k,\vec s_j\times (  \vec s_k \times \vec s_j))=0\\
 \end{array}
\end{equation}
because $(\vec s_k,\vec s_k \times s_1)=0.$
Or, \textbf{ finally}
  \begin{equation}\label{sys}
\begin{array}{c}
 \square \phi_k  + 4\epsilon^2C_{0k} \phi_k(\partial_{0}^{-1}\partial_{1}\phi_1)^2  +
	   4\epsilon^2C_{1kj}\phi_k\phi_j\phi_j=0,\\
 \end{array}
\end{equation}
where
$$C_{0k}=(\vec s_k,\vec s_1\times (   \vec s_k \times \vec s_1))/(\vec s_k,\vec s_k) , C_{1kj}=(\vec s_k,\vec s_j\times (  \vec s_k \times \vec s_j))/(\vec s_k,\vec s_k)$$
or, for normalized $\vec s_k$,
$$C_{0k}= 1-(\vec s_k,\vec s_1)^2 , C_{1kj}=(\vec s_k,\vec s_k-\vec s_j(\vec s_j,\vec s_k))=1-(\vec s_j,\vec s_k)^2. $$
Note also that
$$
C_{02}=1-(\vec s_2,\vec s_1)^2=C_{121},\, C_{03}=1-(\vec s_3,\vec s_1)^2=C_{131}.
$$
Plugging it in \eqref{sys} gives
  \begin{equation}\label{sys}
\begin{array}{c}
 \square \phi_1  +  
	   4\epsilon^2(1-(\vec s_j,\vec s_1)^2)\phi_1\phi_j\phi_j=0,\\
 \square \phi_2  + 4\epsilon^2C_{02} \phi_2(\partial_{0}^{-1}\partial_{1}\phi_1)^2  +
	   4\epsilon^2C_{12j}\phi_2\phi_j\phi_j=0,\\
 \square \phi_3  + 4\epsilon^2C_{03} \phi_3(\partial_{0}^{-1}\partial_{1}\phi_1)^2  +
	   4\epsilon^2C_{13j}\phi_k\phi_j\phi_j=0, 
 \end{array}
\end{equation}
where
$$
C_{02}=1-(\vec s_2,\vec s_1)^2=C_{121},\, C_{03}=1-(\vec s_3,\vec s_1)^2=C_{131}.
$$
For orthonormal vectors $(s_i,s_k)=\delta_{ik}$, one have
\begin{equation}\label{ort}
C_{0k}= 1-\delta_{1k} , C_{1kj}= 1-\delta_{jk},
\end{equation}
that yields
  \begin{equation}\label{sysort}
 \square \phi_k  + 4\epsilon^2( 1-\delta_{1k}) \phi_k(\partial_{0}^{-1}\partial_{1}\phi_1)^2  +
	   4\epsilon^2(1-\delta_{jk})\phi_k\phi_j\phi_j=0, 
\end{equation}
or, expanding 
  \begin{equation}\label{sysort}
 \square \phi_k  + 4\epsilon^2( 1-\delta_{1k}) \phi_k(\partial_{0}^{-1}\partial_{1}\phi_1)^2  +
	   4\epsilon^2 \phi_k\phi_j\phi_j- 4\epsilon^2 \phi_k^3=0.
\end{equation}
The system reads
\begin{equation}\label{sysort1}
\begin{array}{c}
\square \phi_1  + 
	   4\epsilon^2 \phi_1(\phi_2^2+\phi_3^2)=0,\\
\square \phi_2  + 4\epsilon^2\phi_2(\partial_{0}^{-1}\partial_{1}\phi_1)^2  +
	   4\epsilon^2 \phi_2(\phi_1^2+  \phi_3^2)=0,\\
 \square \phi_3  + 4\epsilon^2  \phi_3(\partial_{0}^{-1}\partial_{1}\phi_1)^2  +
	   4\epsilon^2 \phi_3(\phi_1^2+\phi_2^2)=0.
 \end{array}
\end{equation}
A choice of $\phi_1=0,$ gives
\begin{equation}\label{sysort2}
\begin{array}{c}
\square \phi_2  + 
	   4\epsilon^2 \phi_2\phi_3^2=0,\\
 \square \phi_3  + 
	   4\epsilon^2 \phi_3\phi_2^2=0.
 \end{array}
\end{equation}

The minimal choice in (\ref{red}) is
$$
\phi_1=0,\phi_2=\phi_3=\phi.
$$
It is the superposition in spin and isospin states. Then, for the $k=1$ we obtain zero identity, for, $k=2,3$ we have the  same equations of known $\phi^4$ model   with zero mass.
 \begin{equation}\label{fin}
\begin{array}{c}
 \square \phi  + 
	   4\epsilon^2\phi^3=0.
 \end{array}
\end{equation}
It is the case that is maximally close to the Nahm one, but in 1+1.

\section{Towards a solution. }
\label{DW}
\subsection{Projecting technique application}
Consider an equation
\begin{equation}
 \square \phi=F(\phi),
	\end{equation}
	for arbitrary dependence in the r.h.s.. Denoting
\begin{equation}
\begin{array}{c}
  \phi_t=u,  
  \phi_x=v,
	 \end{array}
\end{equation}
 gives the system
\begin{equation}\label{ev}
\begin{array}{c}
 u_t-v_x=F(\int_{-\infty}^xv(y)dy),\\   
  v_t - u_x=0.
	 \end{array}
\end{equation}
The projectors 
\begin{equation}\label{P}
   P_{\pm}=\frac{1}{2}\left(
       \begin{array}{cc}
       1&  \pm 1   )\\
        \pm 1  &  1
       \end{array}
     \right)
\end{equation}
split the linearized system \eqref{ev} in d'Alembert manner
The identity
\begin{equation}
	(P_1+P_2)\psi=\psi
\end{equation}
reads as 
  transformation of fields and its inverse.
\begin{equation}
\begin{array}{c}
	\Pi= \frac{1}{2}( u+   v)\\
\Lambda	=\frac{1}{2}( u-   v)
	 \end{array}
\end{equation}
Acting by the projectors on the evolution system (\ref{ev})  yields
 \begin{equation}
\begin{array}{c}
 \Pi_t-\Pi_x=\frac{1}{2}F(\int_{-\infty}^x(\Pi -\Lambda)dy),\\   
  \Lambda_t + \Lambda_x=\frac{1}{2}F(\int_{-\infty}^x(\Pi -\Lambda)dy),
	 \end{array}
\end{equation}
that describes interaction of essentially one-dimensional waves - gives a \textbf{next step to the Nahm model}.
Asymptotically, for a localized in space solutions, otherwise for a specified initial data
$\Lambda=0$ we have
\begin{equation}
\begin{array}{c}\label{pi}
 \Pi_t-\Pi_x=\Pi_{\xi}=\frac{1}{2}F(\int_{-\infty}^x(\Pi)dy),\\   
  \pi_{\xi\xi}=\frac{1}{2}F(\pi)),
	 \end{array}
\end{equation}
if $\Pi=\pi_{\xi},\, \xi=\frac{1}{2}(x-t),\,\eta=\frac{1}{2}(x+t)$. 

In the case of the Eq. \eqref{fin1} it looks as nonlinear cubic oscillator
\begin{equation}\label{finS}
  \pi_{\xi\xi}=4\epsilon^2 \pi^3.
	 \end{equation}
The equation \eqref{finS} has elliptic solutions \cite{Cor}, see details in the Sec. \ref{Dr}.
\subsection{A path to wavetrains as eventually particles wavefunctions}
Just remind that the anzatz with $\beta<<1,\beta x=x',\beta t=t'$,
\begin{equation}\label{fin2}
  \phi=A(\beta x,\beta t)\exp i[kx-\omega t]+c.c. ,
	 \end{equation}
after plugging in \eqref{fin} and holding nonlinear resonance terms (e.g. \cite{L}) in the first order by the small parameter $\beta$ yields, taking into account the dispersion relation $\omega=\pm k$ and with the rule
 \begin{equation}\label{fin1}
   A_t =\beta A_{t'} . 
	 \end{equation}
It leads to the integrable (in fact - ordinary) equation
 \begin{equation} \label{finE1}
   A_t -A_{x}=\frac{6\epsilon^2}{i k}A^*A^2,. 
	 \end{equation}
that could be solved in terms of elliptic functions.

In the case of \eqref{sysort2} one can obtain approximate solution by the
similar anzatz with $\beta<<1$,
\begin{equation}\label{fin2}
  \phi=A(\beta x,\beta t)\exp i[kx-\omega t]+c.c. ,\phi_3=B(\beta x,\beta t)\exp i[kx-\omega t].
	 \end{equation}
The same manipulations in the first order by the small parameter $\beta$ yields
 \begin{equation}\label{finE2}\begin{array}{c}
 A_t -A_{x}=\frac{6\epsilon^2}{i k}(A^*B^2 +B^*A^2),\\
B_t -B_{x}=\frac{6\epsilon^2}{i k}(B^*A^2 +A^*B^2).
  \end{array} 
	 \end{equation}
It is also solvable as a  system of ODE.

\section{Lagrange density reductions}. 
\label{L}

The Lagrangian density is equal to  (see, e.g. \cite{FS}), 
\begin{equation}\label{Lagr}
\mathcal L=\frac{1}{4}\vec T_{\mu\nu}\vec T^{\mu\nu}=\frac{1}{2}(\vec T_{0i}\vec T^{0i}+\frac{1}{2}\vec T_{ik}\vec T^{ik})=\frac{1}{2 }(\frac{1}{2}\vec T_{ik}\vec T_{ik}-\vec T_{0i}\vec T_{0i}),
\end{equation}
the fields are normalized as in \cite{KP}.
The definition \eqref{Tmn} of the tensor $\vec T_{\mu\nu}$ components with account for Lorentz gauge \eqref{Lor} 
 \begin{equation}\label{Lor1}
\vec A_{0}= \partial_{0}^{-1}\partial_{k}\vec A_{k},
\end{equation}
gives
 \begin{equation}\label{vecTmn}
\vec T_{\mu\nu} =\vec A_{\nu,\mu} - \vec A_{\mu,\nu}- 2\epsilon [\vec A_{\mu}\times \vec A_{\nu}],
\end{equation}
see again \cite{KP}. The time-space components of the tensor are
 \begin{equation}\label{T0i}
\vec T_{0i} =\partial_{0}^{-1}\partial_{k}\frac{\partial \vec A_{k}}{\partial x_i} - \frac{\partial \vec A_i}{\partial t}- 2\epsilon [(\partial_{0}^{-1}\partial_{k}\vec A_{k})\times \vec A_{i}],
\end{equation}
the sum by $k$ is implied. The 3D subtensor looks as \eqref{vecTmn}. The reduction (\ref{red}) reads
 \begin{equation}\label{T0ired}
\vec T_{0i} =\partial_{0}^{-1}\partial_{k}\frac{\partial   \phi_{k}}{\partial x_i}\vec s_k  - \frac{\partial \phi_i}{\partial t}\vec s_i- 2\epsilon (\partial_{0}^{-1}\partial_{k} \phi_{k})\phi_{i}[\vec s_k\times \vec s_i ],
\end{equation}
and
\begin{equation}\label{vecTik}
T_{ik}=\frac{\partial \phi_i(x)}{\partial x_k} \vec s_i -\frac{\partial \phi_k(x)}{\partial x_i}\vec s_k
- 2\epsilon  \phi_i(x)\phi_k[\vec s_i\times \vec s_k ].  
\end{equation}

Its   1+1   space version for the SU(2) gauge (compare with \cite{L1}) gives
   \begin{equation}\label{T0ired}
\vec T_{0i} =\partial_{0}^{-1} \frac{\partial^2   \phi_{1}}{\partial x^2}\vec s_1\delta_{i1}  - \frac{\partial \phi_i}{\partial t}\vec s_i- 2\epsilon (\partial_{0}^{-1}\frac{\partial   \phi_{1}}{\partial x})\phi_{i}[\vec s_1\times \vec s_i ],
\end{equation}
and
\begin{equation}\label{vecTik}
\vec T_{ik}=\frac{\partial \phi_i }{\partial x}\delta_{k1} \vec s_i -\frac{\partial \phi_k }{\partial x}\delta_{i1}\vec s_k
- 2\epsilon  \phi_i \phi_k[\vec s_i\times \vec s_k ].  
\end{equation}
Then
\begin{equation}\label{Lag1}
\begin{array}{c}
\vec T_{0i} \vec T_{0i}=( \partial_{0}^{-1} \frac{\partial^2   \phi_{1}}{\partial x^2}\vec s_1\delta_{i1}  - \frac{\partial \phi_i}{\partial t}\vec s_i- 2\epsilon (\partial_{0}^{-1}\frac{\partial   \phi_{1}}{\partial x})\phi_{i}[\vec s_1\times \vec s_i ])\\( \partial_{0}^{-1} \frac{\partial^2   \phi_{1}}{\partial x^2}\vec s_1\delta_{i1}  - \frac{\partial \phi_i}{\partial t}\vec s_i- 2\epsilon (\partial_{0}^{-1}\frac{\partial   \phi_{1}}{\partial x})\phi_{i}[\vec s_1\times \vec s_i ]).
\end{array}
\end{equation}
in the case of $\phi_1=0,$
\begin{equation}\label{Lag1}
\begin{array}{c}
\vec T_{0i} \vec T_{0i}= \frac{\partial \phi_i}{\partial t} \frac{\partial \phi_i}{\partial t}(    \vec s_i   \cdot \vec s_i ).
\end{array}
\end{equation}
Similarily
\begin{equation}\label{vecTik}\begin{array}{c}
\vec T_{ik}\vec T_{ik}=\frac{\partial \phi_i }{\partial x}\vec s_i  (\frac{\partial \phi_i }{\partial x} \vec s_i -\frac{\partial \phi_1 }{\partial x}\delta_{i1}\vec s_1
- 2\epsilon  \phi_i \phi_1[\vec s_i\times \vec s_1 ]) \\-
\frac{\partial \phi_k }{\partial x} \vec s_k(\frac{\partial \phi_1 }{\partial x}\delta_{k1} \vec s_1 -\frac{\partial \phi_k }{\partial x} \vec s_k
- 2\epsilon  \phi_1 \phi_k[\vec s_1\times \vec s_k ])-\\
( 2\epsilon  \phi_i \phi_k[\vec s_i\times \vec s_k ]) (\frac{\partial \phi_i }{\partial x}\delta_{k1} \vec s_i -\frac{\partial \phi_k }{\partial x}\delta_{i1}\vec s_k
- 2\epsilon  \phi_i \phi_k[\vec s_i\times \vec s_k ]). 
 \end{array}
\end{equation}
and, for normalized $\vec s_i$,
\begin{equation}\label{vecTik}\begin{array}{c}
\vec T_{ik}\vec T_{ik}=\frac{\partial \phi_i }{\partial x} \frac{\partial \phi_i }{\partial x} -\frac{\partial \phi_1 }{\partial x}  \frac{\partial \phi_1 }{\partial x} 
 \\-
\frac{\partial \phi_1 }{\partial x}  \frac{\partial \phi_1 }{\partial x}  +\frac{\partial \phi_k }{\partial x} \frac{\partial \phi_k }{\partial x}  \\
    4\epsilon^2  \phi_i^2 \phi_k^2[\vec s_i\times \vec s_k ][\vec s_i\times \vec s_k ]. 
 \end{array}
\end{equation}
Evaluating, 
$[\vec s_i\times \vec s_k ][\vec s_i\times \vec s_k ]=1-(\vec s_i\cdot\vec s_k)^2=1-\delta_{ik}, $ one arrives at
\begin{equation}\label{vecTik}\begin{array}{c}
\vec T_{ik} \vec T_{ik}=2\sum_{2}^3\frac{\partial \phi_i }{\partial x} \frac{\partial  \phi_i }{\partial x}+  
     4\epsilon^2 ( \phi_i^2 \phi_k^2 -\phi_k^2 \phi_k^2). 
 \end{array}
\end{equation}
 For    the case of $\phi_1=0,$ one have
\begin{equation}\label{vecTik}\begin{array}{c}
\vec T_{ik}\vec T_{ik}=2 (\frac{\partial \phi_2 }{\partial x}^2 +\frac{\partial  \phi_3 }{\partial x}^2 ) 
 + 
   8\epsilon^2 \phi_2^2 \phi_3^2. 
 \end{array}
\end{equation}

Finally, the Lagrange function is
\begin{equation}\label{Lagr1+1}\begin{array}{c}
\mathcal L= -\frac{1}{2} \frac{\partial \phi_i}{\partial t} \frac{\partial \phi_i}{\partial t}   +\\ \frac{1}{2}[  (\frac{\partial \phi_2 }{\partial x})^2 + (\frac{\partial  \phi_3 }{\partial x})^2 ]
 + 
   2\epsilon^2 \phi_2^2 \phi_3^2.  \end{array}
\end{equation}
In the case $\phi_2=\phi_3=\phi$ it is simplified as 
 \begin{equation}\label{Lagr1+1s}
\mathcal  L= - \phi_t^2+ \phi_x^2  +2\epsilon^2\phi^4(x). 
\end{equation}
It is coinside with one of classical $\phi^4$ model case, derived and used, after reduction in \cite{L1} for quasiclassical correction theory. The Euler equation for \eqref{Lagr1+1s} coincides with \eqref{fin}.
 
\section{ Generalized zeta-function regularization of Maslov continual integral}. 
\label{Q}
\subsection{Action integral expansion}
The energy evaluation is based on  calculation of the evolution operator determinant. Its divergence is compensate by  a special choice of the theory basic parameters  using a freedom in the definitions.   
We briefly explain   its origin as well as the small parameter appearance, proportional to  $  \sim \hbar$,  used in the quasiclassical expansion. We keep oursleves in the 1+1 space, the 1+d case is  shown at Appendix.

The approach was presented by Maslov in \cite{Mas}.The action functional on a quantum vector field $\vec \phi=\{\phi_{\alpha},\alpha=2,3\}\in \mathbb{H}$ is defined as integral over space-time stripe $t\in[0,\tau],\overrightarrow{x}\in\mathbb{R}$
\begin{equation}\label{S}
	S(\vec\phi)=\int_0^\tau \int_{\mathbb{R}} \left(\frac{1}{2}\left(\frac{\partial \vec\phi}{\partial t}\right)^2- \frac{1}{2}\left(\frac{\partial\vec \phi}{\partial x}\right)^2-V(\vec\phi)\right)dx dt.
\end{equation}
 We adjust the regularization (renormalization) scheme \cite{L,KwL} to the problem under consideration, having in mind the Lagrange function \eqref{Lagr1+1}.
The regularization consists of two steps. First is based on the assumption, that for  a   vacuum state the corrections should vanish \cite{Mas}.

 Let us expand the action integral 
around a specific classical field $\vec\varphi$ over 1+d space-time.
\begin{equation}\label{vecphi}
	\vec\phi=\vec\varphi+\sum_j w_j \vec \chi_j
\end{equation}
with the appropriate basis $\vec\chi_j$ and approximate \eqref{S} as
\begin{equation}\label{Sexp}
	S(\vec\phi)=S(\vec \varphi)+\sum_{j,k}\frac{\partial^2 S}{\partial w_j \partial w_k}(\vec\varphi)w_j w_k+\dots.
\end{equation}
that,  in turn, defines quasiclassical form of the path integral 
\begin{equation}\label{qc}
	\int_{\mathbb{H}} e^{\frac{i}{\hbar}S(\vec\phi)}D\vec\phi\approx e^{\frac{i}{\hbar}S(\vec\varphi)}\int_{\mathbb{R}} e^{\frac{i}{\hbar}\sum_{j,k}\frac{\partial^2 S}{\partial w_j \partial w_k}(\vec\varphi)w_j w_k}\prod_f dw_f
\end{equation}
with $\vec \varphi$ as the classical path with boundary conditions $\vec \varphi(0,\overrightarrow{x})$ , $\vec \varphi(\tau,x) $ and ${\vec\chi_j}$ as a basis.

Plugging \eqref{vecphi} into \eqref{S},
we obtain for the second  derivative
\begin{equation}
	\frac{\partial^2 S}{\partial w_j \partial w_k}(\vec \varphi)= \int_0^{\tau} \int_{\mathbb{R}^d}  \left(\frac{\partial \vec  \chi_j}{\partial t}\frac{\partial \vec \chi_k}{\partial t}- \frac{\partial\vec  \chi_j}{\partial x}\frac{\partial\vec  \chi_k}{\partial x}-V_{\vec \varphi\vec \varphi}(\vec \varphi)  \vec \chi_j\vec \chi_k\right)d x dt
\end{equation}
where
\begin{equation}\label{V}
V_{\vec \varphi\vec \varphi} \vec \chi_j\vec \chi_k=V_{  \varphi_{\alpha}\varphi_{\beta}}  \chi_{j\alpha}\chi_{k\beta}.\end{equation}

For the basic functions $\vec \chi_k$ from a   Hilbert space $\mathbb{H}$
\begin{equation}
	\frac{\partial^2 S}{\partial w_j \partial w_k}(\vec \varphi)= \int_0^{\tau} \int_{\mathbb{R}}  \left(  \frac{\partial^2 \vec \chi_j}{\partial x ^2}-\frac{\partial^2 \vec \chi_j}{\partial t^2}\vec -V_{\vec \varphi\vec \varphi}(\vec\varphi)\vec\chi_j\right)\vec\chi_kd x  dt \end{equation}

\subsection{Rescaling the integral}

  Let us denote $\tau,\lambda$ as time and space scale parameters, $\epsilon $ is used as interaction parameter. The equations of motion  as \eqref{fin} determine a link between them. 
Introducing dimensionless variables $ x=\lambda x'$, $t= \tau t'$ we rescale as
\begin{equation}
	\frac{\partial^2 S}{\partial w_j \partial w_k}(\vec\varphi)=\frac{\lambda}{\tau} \int_0^1 \int_{\mathbb{R}}  \left( \frac{\tau^2}{\lambda^2}\frac{\partial^2 \vec\chi_j}{\partial x'^2}- \frac{\partial^2\vec \chi_j}{\partial t'^2} -\tau^2V_{\vec \varphi\vec \varphi}(\vec\varphi)\vec\chi_j\right)\vec\chi_kd x' dt'. 
\end{equation}
The factor by the integral defines the quasiclassical expansion parameter, its value being small, allows to cut the expansion at some level. A link between $\lambda$ and $\tau$ is found either from  evolution equation \eqref{finE2} (dispersion relation in classical mechanics) or from realation between momentum, energy  and mass in quantum theory. To be sure that a contribution of the last term is also of order one, we use a link between scale in time $\tau$ and constant of interaction $\epsilon$ that is defined in rather ambiguous way via renormalization procedure (see Sec. \ref{mass}) .

Jumping back into (\ref{qc}) we write the internal factor as 
\begin{equation}
	\int _{\mathbb{H}} e^{\frac{i \lambda^d}{\tau\hbar}\sum_{j,k}\int_0^1 \int_{\mathbb{R}}  \left(\frac{\tau^2}{\lambda^2}  \frac{\partial^2\vec \chi_j}{\partial x'^2} - \frac{\partial^2\vec \chi_j}{\partial t'^2} -\tau^2V_{\vec \varphi\vec \varphi}(\vec\varphi)\vec\chi_j\right)\vec\chi_kd\overrightarrow{x'} dt'w_j w_k}\prod_f dw_f,
\end{equation}
where $V_{\vec \varphi\vec \varphi}(\vec\varphi)$ acts as prescribed by \eqref{V}
\begin{equation}
[V_{\vec \varphi\vec \varphi}(\vec\varphi)\vec\chi_j]_{\alpha}=V_{  \varphi_{\alpha}\varphi_{\beta}}  \chi_{j\beta}.
\end{equation}
In the case of the Lagrangean \eqref{Lagr1+1}  the matrix $V$ in the isotopic subspace
\begin{equation}
\begin{array}{c}
V_{11}=V_{  \varphi_{2}\varphi_{2}}  =4\epsilon^2\phi_3^2,\\
V_{12}=V_{21} =V_{  \varphi_{2}\varphi_{3}}  =4\epsilon^2\phi_2\phi_3,\\
V_{22}=V_{  \varphi_{3}\varphi_{3}}  =4\epsilon^2\phi_2^2.
\end{array}
\end{equation}

Transformations in both spaces $\phi$ and $\chi$ are changing definition of a principal state of the theory. So, if one substitute
\begin{equation}
  \chi_{j\beta}=a^j_{\alpha\beta}\eta^j_{\beta},
\end{equation}
so that
\begin{equation}
 V^j_{\alpha\beta} \eta^j_{\beta}=v(\vec \phi)\eta^j_{\alpha},
\end{equation}
The determinant of the matrix  $V$ is zero, hence eigenvalues are
\begin{equation}\label{v10}
v_0=0,\,v_1=4\epsilon^2(\phi_2^2+\phi_3^2).
\end{equation}
The self-action of the new basic states 
\begin{equation}\begin{array}{c}
 \eta^{J0}_{1}=-\frac{\phi_3}{\phi_2} \eta^{j0}_{2},\eta^{j1}_{1}=-\frac{\phi_2}{\phi_3}\eta^{j1}_{2}.\end{array}
\end{equation}
is defined by correspondent equations that yields in different mass corrections for the principle fields.

\subsection{The final action: spectral zeta function}
\label{zeta}

The second step of the renormalization is following; introduce  a new normalization parameter $ r$ of the basic functions in the Maslov integral construction \cite{Mas}. We can rewrite the integral by introducing a scalar product
\begin{equation}
	(\vec\eta_k,\vec \eta_j)=\int_0^1 \int_{\mathbb{R}}  \vec\eta^*_k\vec \eta_j d x'dt'=r^{-2}\delta_{jk}
\end{equation}
and an operator
\begin{equation}\label{Lform}
	\mathcal L=-\frac{i\lambda }{\pi\hbar r^2\tau}\left(  \frac{\tau^2}{\lambda^2}\frac{\partial^2}{\partial x'^2}- \frac{\partial^2}{\partial t'^2} -\tau^2V_{\vec \varphi\vec \varphi}(\vec\varphi)\right),
\end{equation}
The quasiclassical (Maslov) functional integral \eqref{qc} is written as 
\begin{equation}
e^{\frac{i}{\hbar}S(\vec\varphi)}\int_{\mathbb{H}} e^{-r^2\pi\sum_{j,k}(\vec\eta^k,\mathcal L\vec \eta^j)w_j w_k}\prod_f dw_f
\end{equation}

For the Hermitian $\mathcal L$ the eigen basis chosen yields 
\begin{equation}
	e^{\frac{i}{\hbar}S(\varphi)}\int_{\mathbb{R}} e^{-\pi\sum_j\lambda_j w_j^2}\prod_f dw_f,
\end{equation}
and, after Gauss integrals evaluation,
\begin{equation}
	e^{\frac{i}{\hbar}S(\varphi)}\prod_j (\lambda_j)^{-\frac{1}{2}},
\end{equation}
having in mind that the zero values do not contribute, and the degeneracy of the eigenvalues $\lambda_j$ account, formally,
\begin{equation}
	\frac{e^{\frac{i}{\hbar}S(\varphi)}}{\sqrt{\det[\mathcal L]}}.
\end{equation}

To rewrite the determinants of both operators in a form, which  allow the subtraction, we use a   generalized zeta-function:  
\begin{equation}\label{zet1}
	\zeta_{\mathcal L}(s)=\sum_j \lambda_j^{-s},
\end{equation}
where $\lambda_j$ are nonzero eigenvalues of $\mathcal L$.  
Such definition of the generalized zeta-function should be interpreted as analytic continuation to the complex plane of $s$ from the half plane $,\exists\, \sigma, \, \Re s > \sigma$ in
which the sum converges.
The right side derivative
relation  with respect to $s$ at the point $s=0$
define the determinant
\begin{equation}
\label{lndet}
\ln(\det \mathcal L)=
\zeta_{\mathcal L}^{\prime}(0).
\end{equation}

The generalized zeta-function (\ref{zet1}) admits the
representation via the Green function of the
operator $\partial_{\eta}+\mathcal L$. A link to the  Green function diagonal elements
(\emph{heat kernel formalism}) has been used in quantum theory
since works by Fock (1937) \cite{F}.
The zeta function in 1+1 space is constructed through a set of transformations on the heat equation Green function, the variables $\eta,t,t'$  extended for the whole  axis (for 1+d case see Appendix)
\begin{equation}\label{GreenMain}\begin{array}{c}
	\left(\frac{\partial}{\partial \eta}+\mathcal L\right)g_{\mathcal L}\left(\tau,t,t',x, x'\right)=\\ \delta(\tau)\delta(t-t')\delta\left( x, x')\right).
	\end{array}
\end{equation}
Boundary conditions on the Green function are chosen   the same as for the base  functions in Maslov representation and an additional condition is applied
\begin{equation}
	\forall_{\eta<0}\ g_{\mathcal L}\left(\eta,t,t', x, x'\right)\equiv 0.
\end{equation}
The freedom in a vacuum choice allows to divide $\mathcal L=L+L_0$. In the case of 1+1 space, it yields for \eqref{v10}
\begin{equation}
	L=-\frac{i\lambda}{2\pi\hbar r^2\tau}\left(\frac{\tau^2}{\lambda^2}\frac{\partial^2}{\partial x'^2}-\frac{\partial^2}{\partial t'^2}+4\tau^2\epsilon^2(\phi_2^2+\phi_3^2)-C\right),
\end{equation}
while the function
\begin{equation}
	L_0=-\frac{i\lambda}{2\pi\hbar r^2\tau}\left(\frac{\tau^2}{\lambda^2}\frac{\partial^2}{\partial x'^2}- \frac{\partial^2}{\partial t'^2}-C\right),
\end{equation}
defines the vacuum part, that should be extracted as the first  step of a renormalization. 
Here the constant $C$ depends on the particular classical solutions that form the  potential  $4\tau^2\epsilon^2(\phi_2^2+\phi_3^2)$ minimum value.  
 We can build the renormalized zeta function by the extraction as the first step:
\begin{equation}\label{z}
\begin{array}{c}
	 \zeta(s)=\frac{1}{\Gamma(s)}\int_0^{\infty}\eta^{s-1}\int_0^1\int \left(g_L\left(\eta,t,t, x, x\right) - g_{L_0}\left(\eta,t,t, x, x\right)\right)dxdt d\eta,
	 \end{array}
\end{equation}
while the second step of the renormalization is realized by the special choice of the normalization constant $r$.

\subsection{Extra variables}
\label{Ex}

Working with a 1+d space, the calculations are organized as follows. For construction of the generalized zeta function it is  convenient to use the property  (see appendix for details)
\begin{equation}\label{mult}
	g_{L_a+L_b}=g_{L_a}g_{L_b},
\end{equation}
valid for the operators $L_{a,b}$ dependent on different variables.

It is also useful to introduce an additional function
\begin{equation}\label{gammadef}
	\gamma_{L_a}(\eta)=\int_0^1\int g_{L_a}(\eta,t',t',\vec {x'},\vec{x'}) d\vec{x'}dt',
\end{equation} 
for which   (\ref{mult}) holds as well. Plugging it in \eqref{z}
\begin{equation}\label{z}
\begin{array}{c}
	 \zeta(s)=\frac{1}{\Gamma(s)}\int_0^{\infty}\eta^{s-1}\left(\gamma_L(\eta) -\gamma_{L_0}(\eta)\right)d\eta.
	 \end{array}
\end{equation}
 For  \textbf{one-dimensional classical problem solutions}  
\begin{equation}\label{L1}
	L_1=A\left(\frac{\partial^2}{\partial x'^2_1}-4\lambda^2\epsilon^2(\phi_2^2+\phi_3^2)\right),
\end{equation}
where
\begin{equation}
	 A=-\frac{i\tau}{2\pi \hbar r^2\lambda}
\end{equation}
with the explicit form of the classical problem  solution $\varphi_i(x')$ already specified.
For the first step of renormalization in 1+1 it is enough to  restrict $L_0$ to the space variable only.
\begin{equation}
	L_2=-Ac^2\frac{\partial^2}{\partial t'^2},\,L_{10}=A\frac{\partial^2}{\partial x'^2},
\end{equation}
 where  $c^2=\frac{\lambda^2}{\tau^2}$.
Then the expression \eqref{z} is rewritten as 
\begin{equation}\label{z1}
\begin{array}{c}
	 \zeta(s)=\frac{1}{\Gamma(s)}\int_0^{\infty}\eta^{s-1}\gamma_{L_2}\left(\gamma_{L_1}(\eta) -\gamma_{L_{10}}(\eta)\right)d\eta.
	 \end{array}
\end{equation}
We  transform the Green function $g_{L_1}$ for \eqref{GreenMain},    
then the final form of the spectral zeta function is
\begin{equation}\label{finzet}
\begin{array}{c}
	 \zeta(s)=\frac{1}{\Gamma(s)}\int_0^{\infty}\eta^{s-1}\gamma_{L_2}\left(\gamma_{L_1}(\eta) -\gamma_{L_{10}}(\eta)\right)d\eta,
	 \end{array}
\end{equation}
Integrating in \eqref{gammadef} we derive the approximate  expressions for $\gamma_{L_2},\gamma_{L_{10}}$
\begin{equation}
	\gamma_{L_2}=\sqrt{\frac{1}{4\pi i|A|c^2\eta}},\,\gamma_{L_{12}}=\sqrt{\frac{\lambda}{-4\pi i|A|\eta}}.
\end{equation}
A limit of the spectral zeta derivative when $s\rightarrow 0$ depends on behaviour of $|A|^s, \Gamma(s)$   and its derivatives at this vicinity. It also includes the result of the Meillin transform integration. So, investigation of the behaviour and a choice \textbf{the renormalizing factor $r^2$} of  we left till the explicit evaluation of the ingredients of the \eqref{finzet} will be finalized.

\section{Elliptic solutions of Nahm-like reduced model} 
\subsection{Nahm equation for $\phi=\phi_2=\phi_3$}
\label{Dr}
 
The asymptotic of the solutions of the \eqref{fin} is found via \eqref{finS}. In turn, the equation \eqref{finS} that is a rescaling of the Nahms' one ( \cite{Cor}) that accounts the self-action constant $\epsilon$. Inverse rescaling $\phi=\epsilon \pi/2$ gives
 	\begin{equation}\label{fin2}
		\phi''(z)=2\phi^3.
	\end{equation}
Integrating  (\ref{fin2}) includes a constant of integration ( parameter) $b$
   \begin{equation}\label{Wphi4}
(\phi')^2 = (\phi^2)^2 -b^4.
\end{equation}
It corresponds to the case m=0 of the stationary $\phi^4$ model. Solution of Nahm equation - inversion of the elliptic (lemniscate)
integral 
$\ \phi=bsn(ibz,i).$
\begin{equation}\label{dphi}
   \int_0^\phi \frac{d\phi'}{\sqrt{\phi'^4-b^4}} = \frac{1}{b}\int_{0}%
^{\frac{\phi}{b}\ }\frac{ds}{\sqrt{\left( s^{2}-1\right)  \left(
s^{2}+1\right)  }}=z,\,
\end{equation}
yields the Jacobi $sn$ function with the imaginary module $k=i$;
 the constant $b$ enters the solution as amplitude and space scale   parameter such that $\lambda=b^{-1}$. 
\subsection{Drach equation for the Green function diagonal}
  Take  a Laplace transform $\hat{g}_L(p,x,x_0)$ of the  Green function, defined by \eqref{GreenMain} :
   $G(p,x) = \hat{g}_L(p,x,x)$ is a solution of bilinear equation \cite{L}
\begin{equation}\label{Hermit}
    2GG'' - (G')^2 - 4(u(x)-p)G^2+1=0, 
\end{equation}
Such equation was introduced by J. Drach in a different context \cite{D}.
In a case of reflectionless and finite-gap potentials as ours  u(z) =   $-6b^{2}\left( 1-z\right), z=cn^2(bx;k)$, the equation \eqref{Hermit} is
solved  in
 polynomials  $P,Q$
\begin{equation}\label{PQ}
    G(p,x)=P(p,z)/\sqrt{Q(p)}.
\end{equation}
The form  (\ref{PQ})
yields
\begin{equation}\label{linkPQ}
    b^2(\rho(2PP''-(P')^2)+\rho'PP')-(p+u))P^2+Q=0,
\end{equation}
the primes denote derivatives with respect to z, while
 \begin{equation}\label{rho}
 \begin{array}{c}
                   \rho(z) =         z(1-z)(2-z).   \\
                                           \end{array}
 \end{equation}
The polynomials
\begin{equation}\label{PQD}
\begin{array}{c}
 P=p^2+P_1(z)p+P_2(z),  \\
  Q=p^5+q_4p^4+q_3p^3+q_2p^2+q_1p+q_0, \\
\end{array}
\end{equation}
solve the equation \eqref{linkPQ} if
\begin{equation}\label{splitD}
    \begin{array}{c}
       -2P_1  - u  + q_2 = 0, \\
    -2P_2  - P_1^2  - 2 u P_1 +b^2(2\rho P_1"+\rho'P_1')+q_3=0,    \\
       b^2(\rho (2P_2''+2P_1P_1''-(P_1')^2)+\\
       \rho'(P_2'+P_1P_1'))-2P_1P_2)-u(2P_2+P_1^2)+q_2 =0,  \\
       b^2(2\rho(2P_1''P_2+P_1'P_2'+P_1P_2'')+\\
        \rho'(P_1P_2')+P_1'P_2))-P_2^2 - 2uP_1P_2 + q_1=0,  \\
       b^2(\rho(2P_2P_2''-P_2'^2))+\rho'P_2^2+q_0 =0.\\
     \end{array}
\end{equation}
The arguments in (\ref{splitD}) are omitted.
Plugging $u$ and $\rho$  from \eqref{rho} gives 
  $ q_{4}=0,\quad q_{3}=
-21b^{4}, \quad $ $ q_{2}=
q_{1}=\,108b^{8}, q_{0}= 0, $ hence $P_1(z)=
-3b^2(z-1), \quad P_2=18b^4z^2-36b^4z \cite{L}.$
  \begin{equation}\label{Q}
    Q(p)= p(p+3b^2)(-p+3b^2)(12b^4-p^2)=\prod_{i=1}^{i=5}(p-p_i).
\end{equation}
The polynomial $Q$  simple roots $p_i$ are  ordered for real $b$.  

\subsection{Mass as the correction}
\label{mass}
Let us pick up the expressions determining $\hat{\gamma}_{L_1}(p)$ and $\hat\gamma_{L_0}$, integrating by period:
\begin{equation}\label{hatgammap}
    \hat{\gamma}_{L_1}(p) = \int( p^2
-3b^2(z-1)p+18b^4z^2-36b^4z)dx/2\sqrt{Q}.
   \end{equation}
Denote  complete elliptic
lemniscate integrals as  $K(i)=K, E(i)=E$  and integrating, we have
 \begin{equation}\label{hatgammap'}
\begin{array}{c}
  \hat{\gamma}_{L_1}(p)=  [6b^{4}K +2p^{2}K +  36b^{4}\left(  K   -E\right)  -    3b^{2}p\left(
E  -3K\right) ]     /\sqrt{Q},
\end{array}
   \end{equation}
 that,  plugging the \eqref{Q} and fix the choice of $r$ so that $|A|=1$, 
gives the zeta function \eqref{finzet}  via
\begin{equation}\label{finzet1}
\begin{array}{c}
	 \zeta(s)=\\\frac{1}{\Gamma(s)}\int_0^{\infty}\eta^{s-1}\gamma_{L_2}\int_{l}e^{-p\eta}\left(\frac{6b^{4}K +2p^{2}K +  36b^{4}\left(  K   -E\right)  -    3b^{2}p\left(
E  -3K\right)   }{\sqrt{p(p+3b^2)(-p+3b^2)(12b^4-p^2)}} -\gamma_{L_{10}}\right)dpd\eta,
	 \end{array}
\end{equation}
Finaly, the gauge field particle mass in the quasiclasical approximation is evaluated as the limit
 \begin{equation}
m=\zeta'(0).
\end{equation}
\section{Conclusion} We have considered a nonlinear plane wave of SU(2) YM field (see e.g. \cite{Bas}) in a 1+1 space. The numerical evaluation of the integrals in \eqref{finzet} and the mass will be published elsewhere.  Consideration of the 1+d case in our paper allows to apply multidimension theories in thw spirit of   \cite{L2}.
More generally one can apply the generalized semiclassical Foldy-Wouthuysen transformation as e.g. in \cite{Dobr}. 

Of separate interest there is the special case in the Heisenberg ferromagnet theory \cite{KwL}. It is the easy axis case   when the "mass terms" tends to zero. It corresponds the special choice of the magnetic field $B$ value.

\section{Acknowledgement} Thanks to G. Kwiatkowski for useful discussions.

\section{Appendix. The case of d+1 space}The operator $L$ in 1+d has the form
\begin{equation}
	L=-\frac{i\lambda}{2\pi\hbar r^2\tau}\left(\frac{\tau^2}{\lambda^2}\sum_i^{d}\frac{\partial^2}{\partial x_i'^2}-\frac{\partial^2}{\partial t'^2}+4\tau^2\epsilon^2(\phi_2^2+\phi_3^2)-C\right),
\end{equation}
while the operator
\begin{equation}
	L_0=-\frac{i\lambda}{2\pi\hbar r^2\tau}\left(\frac{\tau^2}{\lambda^2}\sum_i^{d}\frac{\partial^2}{\partial x_i'^2}- \frac{\partial^2}{\partial t'^2}-C\right),
\end{equation}
defines the vacuum state.
To explain \eqref{mult},  take $\phi_{1,j}(x)$ as eigenfunctions of $L_1$ with eigenvalues $\lambda_{1,j}$ and $\phi_{2,n}(y)$ as eigenfunctions of $L_2$ with eigenvalues $\lambda_{2,n}$. Due to the  independence of variables, $\phi_{1,j}(x)\phi_{2,n}(y)$ are eigenfunctions of $L_1+L_2$ with eigenvalues $\lambda_{1,j}+\lambda_{2,n}$.  
\begin{equation}
\begin{array}{c}
	g_{L_1+L_2}=\sum_{j,n}e^{-(\lambda_{1,j}+  \lambda_{2,n})\tau} \frac{\phi_{1,j}(x)\phi_{2,n}(y)\phi_{1,j}(x_0)\phi_{2,n}(y_0)}{(\phi_{1,j}\phi_{2,n},\phi_{1,j}\phi_{2,n})} \Theta(\tau).
	\end{array}
\end{equation}
Considering the scalar product we use, we prove 
\begin{equation}
	(\phi_{1,j}\phi_{2,n},\phi_{1,j}\phi_{2,n})=(\phi_{1,j},\phi_{1,j})_1(\phi_{2,n},\phi_{2,n})_2,
\end{equation}
\begin{equation}\begin{array}{c}
	g_{L_1+L_2}=\\ \sum_j e^{-\lambda_{1,j}\tau}\frac{\phi_{1,j}(x)\phi_{1,j}(x_0)}{(\phi_{1,j},\phi_{1,j})_1} \sum_n e^{-\lambda_{2,n}\tau}\frac{\phi_{2,n}(y)\phi_{2,n}(y_0)}{(\phi_{2,n}.\phi_{2,n})_2}\Theta(\tau),\end{array}
\end{equation}	
For 
\begin{equation}\label{gammadefA}
	\gamma_{L_i}(\tau)=\int_0^1\int g_{L_i}(\tau,t',t',\overrightarrow{x'},\overrightarrow{x'}) d\overrightarrow{x'}dt',
\end{equation} 
  (\ref{mult}) holds as well. Then, if 
\begin{equation}
	L_3=A\sum_{n\neq 1}\frac{\partial^2}{\partial x'^2_n},
\end{equation}
\begin{equation}
	\gamma_{L_3}=-\frac{\lambda}{4\pi A\eta}.
\end{equation}
\end{document}